\documentclass[10pt,aps,pre,showpacs,floats,twocolumn,superscriptaddress,longbibliography]{revtex4-2}

\usepackage{comment}
\DeclareUnicodeCharacter{2009}{\,}
\UseRawInputEncoding

\usepackage{lipsum}
\usepackage{mathrsfs}
\usepackage{bm,amsbsy,amssymb,amsmath}
\usepackage{graphics,graphicx,dcolumn,epic,eepic,float,tabularx}
\usepackage{multirow,rotate,rotating,color}
\usepackage[utf8]{inputenc}
\usepackage{relsize}

\usepackage{xcolor}
\usepackage[utf8]{inputenc}
\definecolor{tuered}{RGB}{214,0,74}
\definecolor{tueblue}{RGB}{0,102,204}

\newcommand{\revisedtext}[1]{\textcolor{black}{#1}}

\usepackage{hyperref}
\usepackage{soul}
\begin{document}

\title{Evaporation-Driven Nanowire Self-Assembly in an Elongated Droplet}

\author{Johannes  Sch\"ottner}
\email{j.schoettner@fz-juelich.de}
\affiliation{Helmholtz Institute Erlangen-N\"urnberg for Renewable Energy (IET-2), Forschungszentrum J\"ulich, Cauerstra{\ss}e 1, 91058 Erlangen, Germany}

\author{Qingguang Xie}
\email{q.xie@fz-juelich.de}
\affiliation{Helmholtz Institute Erlangen-N\"urnberg for Renewable Energy (IET-2), Forschungszentrum J\"ulich, Cauerstra{\ss}e 1, 91058 Erlangen, Germany}

\author{Jens Harting}
\email{j.harting@fz-juelich.de}
\affiliation{Helmholtz Institute Erlangen-N\"urnberg for Renewable Energy (IET-2), Forschungszentrum J\"ulich, Cauerstra{\ss}e 1, 91058 Erlangen, Germany}
\affiliation{Department of Chemical and Biological Engineering and Department of Physics, Friedrich-Alexander-Universit\"at Erlangen-N\"urnberg, Cauerstra{\ss}e 1, 91058 Erlangen, Germany}

\date{\today}

\begin{abstract}

\vspace{10px}

Drying of nanowire-laden elongated droplets is a ubiquitous process in printed electronics fabrication, where the resulting deposition pattern critically determines device performance by controlling nanowire alignment, connectivity, and percolating charge-transport pathways. However, the physical understanding of evaporation-driven deposition is still largely derived from studies of spherical droplets on homogeneous substrates. This gap limits the ability to predict and control deposit morphology in realistic printing scenarios. Here, we use mesoscale lattice Boltzmann simulations to investigate the drying of nanowire-laden elongated droplets on wettability-patterned substrates, focusing on the effects of droplet geometry, nanowire interactions, and nanowire length.
The elongated droplet geometry is found to intrinsically induce distinct axial and transverse inhomogeneities in the final deposit.
%\revisedtext{The elongated droplet geometry intrinsically leads to distinct deposition behavior along the long and short directions of the droplet, resulting in spatial inhomogeneities in the final deposit.}
Increasing the effective attraction between nanowires, which mimics changes in surface chemistry or solvent conditions, can improve electrical connectivity but also promotes clustering and local ordering, reducing structural uniformity. In contrast, increasing nanowire length yields a dual benefit by improving long-range connectivity while simultaneously enhancing deposit homogeneity. Our findings provide design guidance for balancing electrical transport and structural uniformity in evaporation-driven printed electronics.
\end{abstract}

\maketitle

Printing and coating are widely used to fabricate next-generation electronic devices, such as flexible displays, wearable sensors, and thin-film solar cells~\cite{steinberger_2024,huang2019}. In printing (e.g., inkjet, aerosol-jet) and coating (e.g., blade coating), conductive inks are routinely deposited as short lines or elongated footprints that evaporate into continuous, low-resistance pathways required for device functionality~\cite{Guyll2025}.
The resulting morphology---and thus the electrical performance---of these printed features is mainly governed by evaporation-driven flows, contact-line dynamics, and particle properties, \revisedtext{with elongated footprints additionally introducing direction-dependent drying dynamics.} 

The physical understanding of evaporation-induced deposition is still largely derived from the case of a single sessile droplet on a homogeneous substrate. The most prominent example is the coffee-ring effect~\cite{Deegan1997}, in which capillary flow toward a pinned contact line transports suspended material to the droplet edge.
Subsequent studies have shown that the resulting patterns emerge from the interplay of capillary flow, Marangoni stresses, contact-line motion, and particle-substrate interactions\revisedtext{~\cite{Gelderblom2022, Wilson2023, anyfantakis2015,DAmbrosio2025}}.
While this framework successfully explains ring-like deposits in circular droplets, its direct extension to elongated or line-like geometries is not straightforward because the flow field, evaporation flux distribution, and geometric confinement differ markedly.

Previous experimental and numerical work examined liquids on patterned substrates~\cite{Wang2019}, evaporation in elongated droplets~\cite{Aboubakri2022}, and film-wise colloidal assembly on wettable or patterned substrates~\cite{xie2025, mino2022, hartmann2019, Kabi2020, Dinh2016}. These studies show that substrate wettability and geometric confinement strongly influence drying-induced transport and deposition. In particular, patterned or heterogeneous substrates can induce film rupture, droplet splitting, capillary bridges, or anisotropic internal flows, thereby redirecting suspended material toward selected regions of the footprint and promoting ordered monolayers, clustered deposits, or edge-aligned structures, rather than simple isotropic or ring-like patterns.

\revisedtext{Most studies of drying sessile droplets focus on droplets with circular footprints. However, elongated or line-like liquid footprints frequently occur in printing processes, particularly when functional inks are deposited as conductive tracks in printed electronics~\cite{Dinh2016,Goh2019}. Their anisotropic geometry leads to different evaporation and transport behavior along the long and short directions of the droplet, which can strongly affect the resulting particle deposit.
Additionally, previous studies have predominantly considered} spherical or isotropic colloids~\cite{kumar2021}, sometimes accounting for DLVO-type interactions, i.e., van der Waals attraction competing with electrostatic repulsion~\cite{Whitby2020, Bhardwaj2010}. Although filamentary particles (e.g., metallic nanowires and carbon nanotubes) can improve connectivity in the final deposit, they remain much less explored, particularly regarding inter-filament interactions and the link between deposition morphology and material functionality~\cite{Tu2018, Belgardt2013, Park2024}. Moreover, the collective network formation governing percolation and electrical transport in deposited networks is rarely addressed~\cite{Xia2024, ye2019}.

In this work, we numerically study the self-assembly of nanowires, modeled as flexible filaments, in evaporating, elongated droplets on chemically patterned substrates.
Drying begins with \revisedtext{contraction mainly along the long direction} toward a near-spherical shape, followed by radial inward recession, resulting in pronounced \revisedtext{spatial inhomogeneities} in the final deposit. By representing the resulting structures with an effective resistor-network model, we link drying dynamics and filament properties to the electrical transport and structural uniformity of the deposited network in evaporation-driven printed electronics. We find that stronger filament interactions improve electrical transport by increasing network connectivity, but also promote clustering and structural heterogeneity. In contrast, longer filaments enhance long-range connectivity while simultaneously improving spatial homogeneity by bridging gaps and suppressing aggregation.

\section{Methods}
\subsection{Fluid-filament simulation method}
We employ the multi-component lattice Boltzmann color-gradient (CG) method~\cite{Gunstensen1991} to model fluid dynamics.
The method achieves phase segregation through a color-blind collision operator combined with a recoloring step~\cite{Leclaire2017}. In the following, we summarize the core elements of the numerical scheme and refer to our earlier publications for a detailed derivation and validation~\cite{Nath2025,xie2025,Xie2018,Schoettner2026}.
The same framework is adopted here with minor adjustments, and the relevant elements are summarized below.

The evolution of the distribution functions $f_i^{k}$ for component $k$ is governed by the lattice Boltzmann equation
\begin{equation}
f_i^{k}(\mathbf{x}+\mathbf{c}_i \Delta t, t+\Delta t)=f_i^{k}(\mathbf{x},t)-\frac{1}{\tau_k}\left(f_i^{k} - f_i^{k,eq}\right),
\end{equation}
where $\mathbf{x}$ is the lattice position, $t$ is time, $\mathbf{c}_i$ are the discrete lattice velocities, $\Delta t$ is the time step, and $\tau_k$ is the relaxation time of component $k$. The equilibrium distribution is defined as
\begin{equation}
f_i^{k,eq}=w_i \rho_k\left[1+\frac{\mathbf{c}_i \cdot \mathbf{u}}{c_s^2}+\frac{(\mathbf{c}_i \cdot \mathbf{u})^2}{2c_s^4}-\frac{\mathbf{u}^2}{2c_s^2}\right].
\end{equation}
where $w_i$ are the lattice weights, $\rho_k$ is the density of component $k$, $\mathbf{u}$ is the fluid velocity, and $c_s$ is the lattice speed of sound. The macroscopic density and momentum fields are computed via
\begin{equation}
\rho_k = \rho_0\sum_i f_i^{k},\qquad\rho \mathbf{u} = \sum_k \sum_i f_i^{k} \mathbf{c}_i,
\end{equation}
with $\rho_0$ the reference density and the total density given by $\rho=\sum_k \rho_k$.

Within the color-gradient framework, the collision procedure consists of a color-blind relaxation step followed by a recoloring operation that enforces phase segregation and maintains a sharp interface. In the present implementation, interfacial tension is incorporated through a continuum surface force (CSF) formulation \cite{Brackbill1992, Gu2023, Beunen2026}, rather than through a classical population-level perturbation step. Specifically, the interface curvature is computed from gradients of the color field $C=(\rho_1-\rho_2)/(\rho_1+\rho_2)$, and the resulting capillary force is applied as a body force prior to the collision step, weighted by the local density fractions. This formulation improves numerical stability and significantly reduces spurious currents compared with perturbation-based approaches.

Substrate wetting is modeled using the geometry-aware formulation of Akai~\cite{Akai2018}. In this approach, wall normals are reconstructed from the voxelized surface geometry, and a prescribed contact angle $\theta$ is imposed by consistently modifying the interfacial indicator in the solid ghost layer.

\begin{figure}[htbp]
    \centering
    \includegraphics[width=1.0\linewidth]{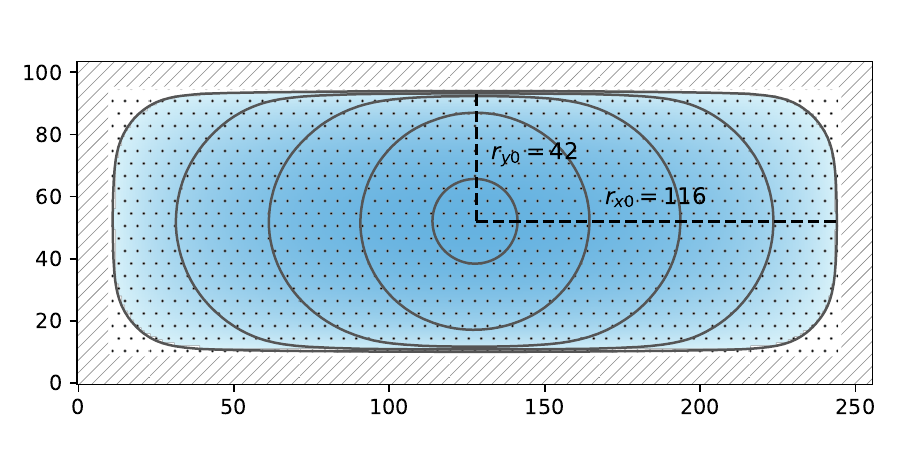}
    \caption{Top view of an elongated droplet on a patterned substrate with dimensions $r_{y0}=42$ and $r_{x0}=116$. Contour lines indicate the contact line at selected evaporation times. Blue shading shows the initial droplet height profile ($h\approx30$). The dotted patch region beneath the initial droplet corresponds to a receding contact angle $\theta_r=30^\circ$, whereas the surrounding substrate corresponds to $\theta_a=90^\circ$.}
    \label{fig:illustration1}
\end{figure}

\textit{Evaporation} is modeled as an interfacial mass sink that converts liquid resting populations into gas~\cite{Nath2025}. For numerical stability, the local evaporation rate is constrained such that the extracted mass does not exceed the available resting population at a lattice node. Because evaporation is not intrinsically captured by the color-gradient method, an interfacial evaporative flux must be prescribed. This is possible for the droplet geometry considered, since analytical or approximate expressions for the evaporative flux are available. \revisedtext{In the diffusion-limited regime, the vapor concentration in the gas phase satisfies $\nabla^2 c=0$, with the concentration set by the saturation value at the liquid--gas interface and by the ambient vapor concentration far from the droplet. The resulting evaporation flux therefore generally depends on the complete droplet geometry and surrounding vapor field~\cite{Gelderblom2022,Larsson2023}.} For thin droplets, the diffusion-limited evaporative flux can be written to first order as~\cite{Wray_Moore_2023}
\begin{equation}
J(r,\varphi) = \frac{2}{\pi}\frac{a}{\sqrt{a^2-r^2}}\left[1 + \epsilon\, f_1(r;n)\cos(n\varphi)\right].
\end{equation}
\revisedtext{Here, $r$ is the dimensionless radial coordinate, $\varphi$ the angular coordinate, and $a=a(\varphi)=1+\epsilon\cos(n\varphi)$ describes the contact-line shape, with $\epsilon$ quantifying deviations from circular symmetry.} For the geometry considered (see Fig.~\ref{fig:illustration1}), the initial rectangular wetting patch has half-lengths \(r_{x0}=116\) and \(r_{y0}=42\) lattice units. The corresponding deformation parameter is
\begin{equation}
\epsilon=\frac{r_{x0}-r_{y0}}{r_{x0}+r_{y0}}\approx 0.468.
\end{equation}
The mode-dependent correction function is
\begin{equation}
f_1(r;n)=
r^2\,\frac{1-r^{\,n-2}}{1-r^2}.
\end{equation}
An elliptic footprint is represented by the mode $n=2$. For this mode, the \revisedtext{explicit} first-order correction function vanishes, $f_1(r;2)=0$. 
\revisedtext{However, the elliptic shape remains included through $a(\varphi)=1+\epsilon\cos(2\varphi)$ in the leading term $a/\sqrt{a^2-r^2}$. We note that additional geometric corrections to the evaporation flux arise at second order in $\epsilon$~\cite{Wray_Moore_2023}. For the geometry considered here, these corrections amount to approximately $6.4\%$ of the leading-order term and are therefore negligible.
In the simulations, the spatial variation of evaporation flux is approximated using a local height-dependent flux~\cite{MURISIC2011}, with the same functional form as the non-equilibrium one-sided (NEOS) model~\cite{Larsson2023},}
\begin{equation}\label{EvapFlux}
  J \approx \frac{J_0}{K + \tilde{h}},
\end{equation}
where \(\tilde{h}\) is the droplet height normalized by its initial value.
\revisedtext{The parameter $K$ provides a measure of kinetic effects and the volatility of the liquid and can be interpreted as the inverse of an effective mass-transfer coefficient, while $J_0=2.5\times10^{-5}$ sets the characteristic evaporation rate.
The parameter $K$, chosen to be $K=0.01$, mainly affects the evaporation flux close to the contact line, where the normalized film height $\tilde h$ becomes comparable to $K$; for $\tilde h\gg K$, its influence is small.
To assess the sensitivity to this choice, we additionally performed a pure-droplet simulation with $K=0.001$. Reducing $K$ by a factor of ten increases the maximum velocity magnitude by about $15\%$, while the liquid-volume difference is of similar magnitude. Thus, although decreasing $K$ increases the maximum local evaporation flux, an order-of-magnitude change in $K$ produces only a moderate change in the resolved droplet-scale dynamics.
}
For small values of $K$, Eq.~\ref{EvapFlux} \revisedtext{qualitatively reproduces the increase of the diffusion-limited evaporative flux near the contact line}, while larger values of $K$ correspond to \revisedtext{a stronger limitation of evaporation by the finite rate of mass transfer across the liquid--vapor interface}.
\revisedtext{The present height-dependent approximation is intended for thin, quasi-static droplets with dilute particle loading and provides a qualitative representation of the enhanced evaporation toward the contact line. It does not resolve the full vapor-diffusion field, including higher-order geometric effects or particle-induced changes of the evaporation flux.}

\textit{Filaments}, representing nanowires, are modeled as discrete bead--spring chains.
Their intra-filament connectivity is enforced using finitely extensible nonlinear elastic (FENE) bonds,
\begin{equation}
    U_{\mathrm{FENE}}(d) = -\frac{1}{2} k_s R_m^2 \ln\left(1-\frac{d^2}{R_m^2}\right), \qquad d<R_m ,
\end{equation}
where $d$ is the distance between neighboring beads, $k_s=0.3$ is the spring constant, and $R_m=2.4$ is the maximum bond extension. Excluded-volume interactions are modeled by a Weeks-Chandler-Anderson (WCA)-like potential,
\begin{equation}
    U_{\text{WCA}}(d) = \varepsilon \left[ \left( \frac{d_0}{d} \right)^{12} - 2 \left( \frac{d_0}{d} \right)^6 \right] + \varepsilon, \quad d \leq d_0
\end{equation}
with bead diameter $d_0=2$ and $\varepsilon=0.03$. Stiffness is included by adding a harmonic potential between second-nearest neighbors,
\begin{equation}
    U_{\mathrm{bend}}(d) = \frac{1}{2} k_b (d-2d_0)^2 ,
\end{equation}
where $k_b=8.0$ penalizes deviations from a straight filament backbone.

\revisedtext{For a representative local bending angle of $\alpha=30^\circ$, we estimate an effective bending rigidity by matching the discrete bending energy to $U_{\mathrm{bend}}=B_{\mathrm{eff}}\alpha^2/(2d_0)$, with $d=2d_0\cos(\alpha/2)$, which gives $B_{\mathrm{eff}}\approx1.1$ in lattice units. The capillary-to-bending ratio $\Pi_\gamma=\gamma d_0L^2/B_{\mathrm{eff}}$ is approximately $20$ for $L=12$ and $144$ for $L=32$, while the elastoviscous number $\Pi_v=8\pi\mu\dot{\gamma}L^4/(cB_{\mathrm{eff}})$~\cite{Chakrabarti2021}, with $c=2\ln(2L/d_0)-1$ and a characteristic shear rate $\dot{\gamma}\sim1/T_S$, where $T_S=10^5$ LB time steps is the total evaporation time, is approximately $0.4$ and $13$, respectively. While these values depend strongly on filament dimensions and flow conditions, they indicate stronger relative capillary and viscous forcing in our simulations, particularly for the longer filaments.}
Nanowires are two-way coupled to the lattice Boltzmann fluid through the dissipative coupling scheme of Ahlrichs and D\"unweg~\cite{Ahlrichs1999}.
\revisedtext{Each bead experiences a local drag force based on its velocity relative to the interpolated fluid velocity, while the opposite force is transferred back to the fluid. Consequently, wall-induced changes of the resolved fluid velocity affect the bead drag. However, the point-particle coupling does not explicitly resolve short-range lubrication effects at particle--wall separations below the lattice resolution.}
Confinement within the liquid phase is maintained by a solvation force based on local fluid-density gradients~\cite{Sega2013}, which drives beads away from the liquid--gas interface and is calibrated such that beads remain inside the droplet during drying.
\revisedtext{Within the present point-particle description, this confinement prevents particles from leaving the liquid phase during interface recession, but neglects possible particle accumulation or adsorption at the liquid--gas interface, as observed experimentally in evaporating colloidal droplets~\cite{Bruning2020}.} In the coupled simulations, five substeps are performed to update the bead positions and velocities per lattice Boltzmann (LB) time step. This temporal subcycling improves numerical stability by reducing integration errors in the bead trajectories and allows the beads to sample the underlying interaction potential more accurately. The detailed implementation, integration scheme, and validation of the filament model follow our previous work~\cite{Schoettner2026}.

Filament-filament and filament-substrate interactions are modeled by a Lennard-Jones (LJ) potential,
\begin{equation}\label{Eq:LJ}
U_{\mathrm{LJ}}(d)=\varepsilon\left[\left(\frac{d_0}{d}\right)^{12}-2\left(\frac{d_0}{d}\right)^6\right],
\end{equation}
where $d$ denotes the center-to-center distance between beads, $\varepsilon$ represents the potential well depth (interaction strength), and $d_0=2.0$ is the effective bead diameter at which the potential minimum occurs. The purely repulsive limit, corresponding to the WCA potential, is obtained by truncating and shifting the LJ interaction at its minimum. The interaction is therefore cut off at $d=d_0$, where the potential minimum occurs, ensuring that both the potential and the force vanish continuously while retaining only the repulsive branch. Optional cohesion between filaments is introduced via a finite attractive $\varepsilon>0$, which can be tuned to represent different filament materials, ionic strengths, surface functionalizations, or solvent conditions (e.g., electrostatic contributions modulated by pH or salt concentration)~\cite{Whitby2020,Bhardwaj2010}. In addition, a frictional interaction between filaments and the substrate is included to account for the tangential resistance of beads in contact with the substrate. The friction depends on the normal load acting on the beads, which arises from hydrodynamic stresses, filament-filament interactions, and adhesive bead-substrate interactions ($\varepsilon_s = 0.006$), leading to resistance to motion along the substrate.

\textit{Problem Definition and Assumptions: }
The simulations describe isothermal evaporation of a sessile, initially elongated water-based droplet on a patterned substrate (Fig.~\ref{fig:illustration1}), in the viscous regime with negligible inertial effects. \revisedtext{The long $x$ direction of the elongated droplet is referred to as axial, and the short $y$ direction as transverse.} The droplet shape evolution is assumed to be quasi-static, such that the interface relaxes rapidly compared to the evaporation timescale. This corresponds to a regime with small capillary number and small Bond number, where surface tension dominates over viscous deformation and gravity. Additional dynamic shape instabilities (e.g., breakup) are assumed to be negligible for the moderately elongated footprint considered here ($r_{x0}/r_{y0}\approx 2.8$), but may become relevant for more strongly elongated liquid shapes on substrates~\cite{Ghigliotti2013,Dziedzic2019}. Brownian motion is not explicitly included; the model therefore applies to regimes with sufficiently large P\'eclet number, where capillary-flow-driven advection dominates over diffusion.
\revisedtext{Marangoni effects can arise during droplet evaporation due to surface-tension gradients caused by temperature or surfactant concentration variations~\cite{Hu2005,Gelderblom2022}. In the present model, the droplet is assumed to be isothermal and surfactant-free, and Marangoni effects are therefore neglected.} The droplet is assumed to be deposited on a smooth glass or silicon substrate under ambient laboratory conditions, with a receding contact angle in the range $\theta_r = 20$--$30^\circ$~\cite{Aliakbar2025}. \revisedtext{Filaments can become immobilized by substrate friction and interact with the fluid through two-way coupling, but do not modify the prescribed wetting condition; particle-induced contact-line self-pinning is therefore neglected.} Surface tension is assumed constant and corresponds to water--air at room temperature ($\gamma \approx 0.072\,\mathrm{N/m}$).

The simulations are designed to preserve the relevant control parameters, including the initial droplet deformation parameter, the diffusion-limited evaporation profile, the filament-to-droplet length ratio, the filament aspect ratio, the initial filament concentration, and the effective filament–filament and filament–substrate interactions. The simulated filament system approximately represents a broad range of filamentary materials relevant to flexible and printed electronics, including conductive microfibers~\cite{Zheng2021,Ka2025}, nanowires, nanotubes, and conductive polymer nanofibers~\cite{Zhang2017,Dai2002,Bessaire2017}.

\subsection{Structural characterization method}

We quantify the structural organization of the deposited filament network by evaluating filament alignment, spatial material distribution, and electrical properties, thereby linking filament arrangements to conductive behavior. All reported observables are averaged over independent simulation runs to reveal systematic trends.

\textit{Morphological characterization.}
Filament alignment is quantified using a segment-wise nematic order parameter for bonds connecting adjacent beads. The nematic order parameter in bin $b$ is defined as
\begin{equation}
    S_x^b
    =
    \left\langle
    \frac{1}{N_b^*}
    \sum_{k=1}^{N_b^*}
    \left(2\cos^2\zeta_{x,k}-1\right)\right\rangle,
\end{equation}
where $\zeta_{x,k}$ is the angle between segment $k$ and the axial direction, and $N_b^*$ is the number of segments whose midpoints lie inside bin $b$. This captures the local orientational order of filament segments induced by deposition and therefore quantifies the degree of inter-filament alignment within the considered region.

Spatial material distribution is described by the normalized density profile, $g^b$. For each bin $b$, it is defined as
\begin{equation}
    g^b
    =
    \left\langle
    \frac{N_b/A_b}{n_0}\right\rangle,
    \label{eq:normalized_bead_density}
\end{equation}
where $N_b$ is the number of bead centers in bin $b$, $A_b$ is the corresponding bin area, and $n_0$ is the average density of the analysis domain. For equal-sized bins, this is equivalent to normalizing the bead count in each bin by the mean bead count over all bins. Values of $g>1$ indicate local enrichment, while $g<1$ denotes depletion. Axial or transverse profiles of $S_x^b$ and $g^b$ are obtained by averaging over a central strip whose width is half of the patch width and which is oriented perpendicular to the profile direction. For an axial profile, the averaging is along the transverse direction; for a transverse profile, it is along the axial direction. This captures systematic variations along the direction of interest while minimizing edge effects. Thus, $S_x^b$ describes segment-level alignment, while $g^b$ quantifies the relative spatial distribution of deposited filament material. Both variables serve as structural descriptors for subsequent analysis of network conductivity and current pathways.

\begin{figure*}[htbp]
    \centering
    \includegraphics[width=1.0\linewidth]{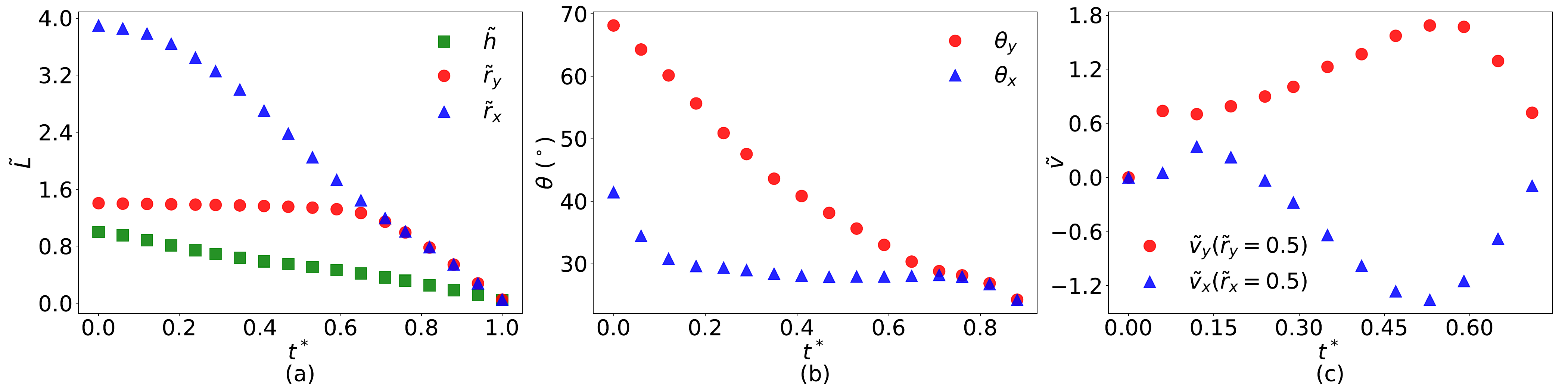}
    \caption{Drying dynamics of a droplet on the patched substrate with a receding contact angle of $30^\circ$, shown as functions of normalized time $t^*$. The contact line remains approximately pinned until $t^* \approx 0.1$, after which the droplet contracts anisotropically in the axial ($x$) direction. At $t^* \approx 0.7$, full depinning occurs and the droplet recedes symmetrically. (a) Normalized droplet height $\tilde{h}$ and footprint radii $\tilde{r}_x$ and $\tilde{r}_y$. (b) Corresponding contact angles $\theta_x$ and $\theta_y$. (c) \revisedtext{Height-averaged} velocity components $\tilde{v}_x$ and $\tilde{v}_y$, evaluated at $\tilde{r}_x=0.5$ and $\tilde{r}_y=0.5$, respectively. \revisedtext{The contact-angle and velocity data are shown only while fluid nodes remain available for evaluation.}}
    \label{fig:result1}
\end{figure*}

\textit{Electrical network model}
To quantify electrical transport in the deposited filament network, the final bead configuration is mapped onto an equivalent resistor network. Each bead is treated as a conducting node, while electrical connections arise from filament segments and sufficiently close bead--bead contacts. Two beads $i$ and $j$ are connected if their Euclidean distance $r_{ij}$ is below a cutoff $r_c = 2.1$. Their corresponding dimensionless conductance is assumed to scale inversely with this distance and is set to $G_{ij} = 1/r_{ij}$, providing a geometric approximation of the conductive pathways formed by the filament network. In this effective model, the conductance $G_{ij}$ represents electrical transport along filament segments as well as between nearby filaments, without explicitly distinguishing between bulk filament conduction and inter-filament contact resistance. A dimensionless reference potential difference $\Delta V = 1$ is applied across the domain along the deposition ($x$) direction. All nodes with $x < x_{\mathrm{src}}$ are assigned to the source electrode, while nodes with $x > x_{\mathrm{snk}}$ define the sink electrode. The node potentials follow from Kirchhoff’s current law and are obtained by numerically solving the resulting sparse linear system. Once the potentials are known, the absolute local current magnitude at node $i$ is evaluated by summing the fluxes through its adjacent edges,
\begin{equation}
    I_i = \frac{1}{2} \sum_j \left| I_{ij} \right|,
\end{equation}
where the edge current is $I_{ij} = G_{ij} ( V_i - V_j )$. This quantity measures the local current throughput of a node rather than the net current, which vanishes for internal nodes by Kirchhoff's current law. The prefactor of $1/2$ accounts for the fact that incoming and outgoing current contributions are both included in the sum of absolute edge currents. To compare different deposits, currents are normalized by a reference current 
\begin{equation}
I_{\mathrm{ref}} = \frac{\varkappa_0 \Delta V A}{L},
\end{equation}
which corresponds to an ideal straight filament of intrinsic material conductivity $\varkappa_0$ spanning the electrode separation $L$ with cross-sectional area $A$. The dimensionless node current is therefore
\begin{equation}
I_i^*=\frac{I_i}{I_{\mathrm{ref}}}.
\end{equation}
Spatial maps of $I_i^*$ reveal the heterogeneity of electrical transport and identify the dominant conductive pathways. The normalization characterizes the relative contribution of individual filaments within the network. The total current \(I_{\mathrm{tot}}\) is obtained from the currents leaving the source electrode. For a fixed electrode placement, we define
\begin{equation}
    \varkappa^*=\left\langle \frac{I_{\mathrm{tot}}}{I_{\mathrm{ref}}} \right\rangle.
\end{equation}
as a dimensionless relative conductivity. Because the deposits are spatially non-uniform and their extent changes with drying conditions and cohesion, $\varkappa^*$ depends on the chosen source--sink geometry and is therefore interpreted as an effective network conductivity for a fixed electrode configuration. The network conductivity calculations were performed in Python using \texttt{NetworkX} for graph construction and connectivity analysis~\cite{Hagberg2008}, and \texttt{SciPy} sparse linear algebra for solving the Kirchhoff equations~\cite{Virtanen2020}.

\section{Results and Discussion}
We begin by characterizing the drying dynamics of the \emph{pure} liquid on the patched substrate, tracking the droplet height, footprint radii, contact angles, and representative capillary-flow velocities. The analysis then turns to the filament-laden droplet, where the coupled drying dynamics lead to redistribution and orientational ordering of the filaments. Finally, we relate the resulting deposit morphologies to electrical transport by analyzing how the relative conductivity of the dried networks depends on filament length and inter-filament cohesion.

\subsection{Evaporation of a pure droplet on a patched substrate}

\begin{figure*}[htbp]
    \centering
    \includegraphics[width=1.0\linewidth]{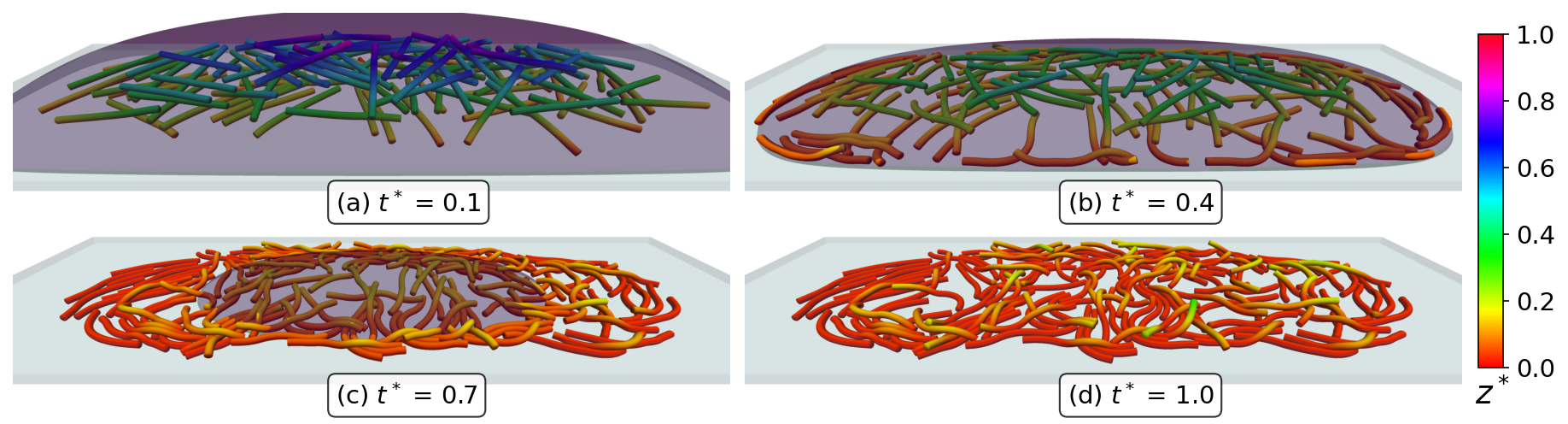}
    \caption{Illustrative rendering of the drying process at different nondimensional times $t^*=t/T_S$ ((a) t*=0.1, ((b) t*=0.4, (c) t*=0.7, and (d) t*=1.0) for non-cohesive filaments of length $L=32$, with an initial concentration of $4\%$ and receding contact angle $\theta_r=20^\circ$. The color bar indicates the normalized vertical coordinate $z^*$. The droplet first contracts axially and later radially after approaching a spherical-cap shape. Capillary flows transport suspended filaments mainly toward the pinned contact lines, while deposited filaments remain immobilized once substrate friction exceeds the solvation force.}
    \label{fig:Snap}
\end{figure*}

We first consider a pure droplet (no filaments) to establish the baseline interfacial dynamics imposed by the line-shaped patch. Fig.~\ref{fig:illustration1} schematically summarizes the droplet geometry, wetting conditions, and the key quantities used to characterize an elongated droplet on a patched substrate. The droplet is initialized with a sharp interface and relaxed to its equilibrium shape on the patterned substrate before evaporation begins. The shaded area of the substrate beneath the initial droplet is assigned a contact angle $\theta_r = 30^\circ$ (defined as receding contact angle), while the remaining surface has a contact angle $\theta_a = 90^\circ$, ensuring that the droplet remains initially pinned and does not spread.  The computational domain has a size of $256 \times 104 \times 36$ lattice units. Prior to drying, its height and footprint radii are approximately $h_0 \approx 30$, $r_{x0} \approx 116$, and $r_{y0} \approx 42$ lattice units.

Fig.~\hyperref[fig:result1]{\ref*{fig:result1}a} shows the temporal evolution of the droplet height and footprint radii in dimensionless form, $\tilde{h}=h/h_0$ and $\tilde{r}_{x,y}=r_{x,y}/h_0$, as functions of the normalized time $t^*=t/T_S$.
%\revisedtext{where $T_S=10^5$ LB time steps is the total evaporation time.
\revisedtext{Velocities are normalized by the characteristic evaporation velocity $v_c=h_0/T_S$, which represents the average rate of droplet-height decrease over the evaporation time.}
The droplet height $\tilde h$ (squares) decreases monotonically throughout the evaporation process, reflecting the continuous loss of liquid volume. In contrast, the footprint radii exhibit distinct behaviors due to substrate pinning and anisotropic contraction. The transverse radius $\tilde r_y$ (circles) initially remains nearly constant, indicating a pinned contact line along the patch edges. At later times it begins to decrease, signaling a depinning transition. The axial radius $\tilde r_x$ (triangles) decreases from the beginning, showing that the droplet contracts preferentially along the axial direction where the contact line is less constrained. The pinning imposed by the patch, together with a low ratio of the capillary relaxation time to the evaporation time scale, leads to an evaporation regime in which the droplet initially contracts predominantly along the axial direction. The resulting anisotropic contraction progressively reduces the footprint eccentricity until the droplet approaches a spherical-cap geometry at $t^*\approx0.65$, corresponding to a height of $\tilde h\approx0.4$ and nearly equal base radii $\tilde r_x\approx\tilde r_y\approx1.5$. After this point the contraction becomes essentially isotropic, and the droplet fully evaporates at $t^*\approx1.0$.

In Fig.~\hyperref[fig:result1]{\ref*{fig:result1}b} the temporal evolution of the contact angles along the transverse ($y$) and axial ($x$) directions is plotted. Initially the axial contact angle is $\theta_x \approx 40^\circ$ and the transversal contact angle is $\theta_y \approx 70^\circ$. During evaporation until $t^*\approx 0.1$ the contact angles are above the receding contact angle $\theta=30^\circ$, but enhanced evaporative flux near the contact line and the associated capillary relaxation lead to a slight contraction of the droplet footprint in the axial direction. As evaporation proceeds, the axial contact angle strongly decreases and eventually reaches the receding value, $\theta_x \le 30^\circ$, at which point depinning occurs and the droplet begins to retract along the axial direction. Over time, continued evaporation and flow-driven relaxation reduce the footprint eccentricity, causing the transverse contact angle to decrease and approach the axial contact angle. They eventually converge as the droplet reaches the shape of a spherical cap and the evaporation continues in the constant contact angle mode.

Fig.~\hyperref[fig:result1]{\ref*{fig:result1}c} shows the temporal evolution of the internal flow velocities evaluated from the center at $\tilde{r}_x=0.5$ in axial and at $\tilde{r}_y=0.5$ in transverse directions. \revisedtext{The corresponding velocity components are height-averaged over the liquid phase.} During the pinned stage, the decreasing contact angle generates a strong outward transverse flow associated with the \textit{Rush-Hour} effect~\cite{Marn2011}. The transverse velocity $\tilde{v}_y$ increases substantially until $t^*=0.55$, exceeding the characteristic evaporative velocity $v_c$ by up to $75\%$. Once transverse depinning occurs, this outward flow weakens. In contrast, the axial velocity $\tilde{v}_x$ initially increases only slightly, reflecting fluid redistribution toward the initially pinned axial contact lines. As evaporation proceeds and the contact angle decreases, \revisedtext{the axial contact line depins, while continued evaporation drives liquid toward the droplet center and the transverse sides.} As a result, the axial flow reverses and $\tilde{v}_x$ gradually becomes negative. Once the droplet approaches a spherical-cap shape and the contact line becomes fully unpinned at $t^*=0.7$, these redistribution flows \revisedtext{are expected to weaken}. Both axial and \revisedtext{transverse} velocities then decay toward zero, indicating that the droplet interior has reached a quasi-steady state with only negligible internal circulation. The same qualitative flow sequence is observed for $\theta_r=20^\circ$.

\subsection{Evaporation of a filament-laden droplet}

Here, we investigate the evaporation dynamics of a droplet laden with suspended filaments. The droplet is initially prepared as a liquid cylinder and allowed to equilibrate on a substrate with patterned wettability, so that it adjusts to the imposed surface-energy landscape. After equilibration, straight, non-overlapping filaments with random positions and orientations are inserted, while ensuring that they remain fully immersed in the liquid and do not intersect the diffuse interface. Evaporation is then initiated by applying the boundary condition with the flux given by Eq.~\ref{EvapFlux}. Because the system operates in the Stokes regime, filaments largely follow the evaporation-driven flow field. In the elongated droplet geometry, capillary-driven flows are stronger in the transverse direction, leading to preferential transport toward the \revisedtext{transverse sides}. As evaporation reduces the droplet volume, the filament concentration increases. In the presence of cohesive filament interactions, this promotes filament agglomeration during drying.

Fig.~\ref{fig:Snap} shows the temporal evolution of such a filament-laden droplet at different times on a rectangular substrate with patterned wettability. A contact angle of $\theta_r = 20^\circ$ is imposed on a rectangular patch beneath the initial droplet, while $\theta_a = 90^\circ$ is assigned to the surrounding substrate. The low contact angle on the patch is chosen as a model value for a strongly wetting printed line and provides a clear wettability contrast to the surrounding substrate. Up to the dimensionless time $t^* \approx 0.1$, the contact line remains pinned in the transverse direction and only partially pinned in the axial direction (Fig.~\hyperref[fig:Snap]{\ref*{fig:Snap}a}). During this stage, evaporation primarily reduces the droplet height, while the footprint shrinks only slightly in the axial direction due to imperfect pinning. Capillary flows begin to transport filaments toward the contact line and align them with the local streamlines. Once filaments reach the contact line, their orientation is governed by the competition between the outward capillary flow and the inward motion of the receding contact line. During pinning, the outward flow dominates, whereas its relative influence decreases strongly once the contact line starts to recede. As a result, filaments near the initial transverse contact line exhibit a pronounced axial alignment, while those near the axial contact line show a mixed orientation with both axial and transverse contributions.

Once the axial contact angle reaches the receding value, the droplet begins to contract freely in the axial direction while remaining pinned transversely. Despite the axial retraction, a weak advection toward the axial edges, indicated by positive $v_x$ in Fig.~\hyperref[fig:result1]{\ref*{fig:result1}c} up to $t^*=0.2$, persists because of the strongly non-uniform evaporative flux following from Eq.~\ref{EvapFlux} further enhancing axial alignment. Filaments anchored at the axial contact lines now experience an additional drag on their free end from the solvation force of the receding contact line, which further enhances the axial alignment (Fig.~\hyperref[fig:Snap]{\ref*{fig:Snap}b}). In the transversal direction filaments continue to be advected toward the contact line, where they initially align with the transverse streamlines and, upon deposition, may further orient along the contact line or along previously deposited filaments. 

As the droplet approaches a spherical-cap geometry, its contraction becomes predominantly radial (Fig.~\hyperref[fig:Snap]{\ref*{fig:Snap}c}). The resulting contact-line motion transports filaments toward the droplet center rather than promoting further alignment along the interface, leading to an increased filament concentration in the central region (Fig.~\hyperref[fig:Snap]{\ref*{fig:Snap}d}) compared with the previous stage (Fig.~\hyperref[fig:Snap]{\ref*{fig:Snap}c}).

\subsubsection{Nematic order and density profile}
\label{subsub1}

\begin{figure}[h]
    \centering
    \includegraphics[width=0.99\linewidth]{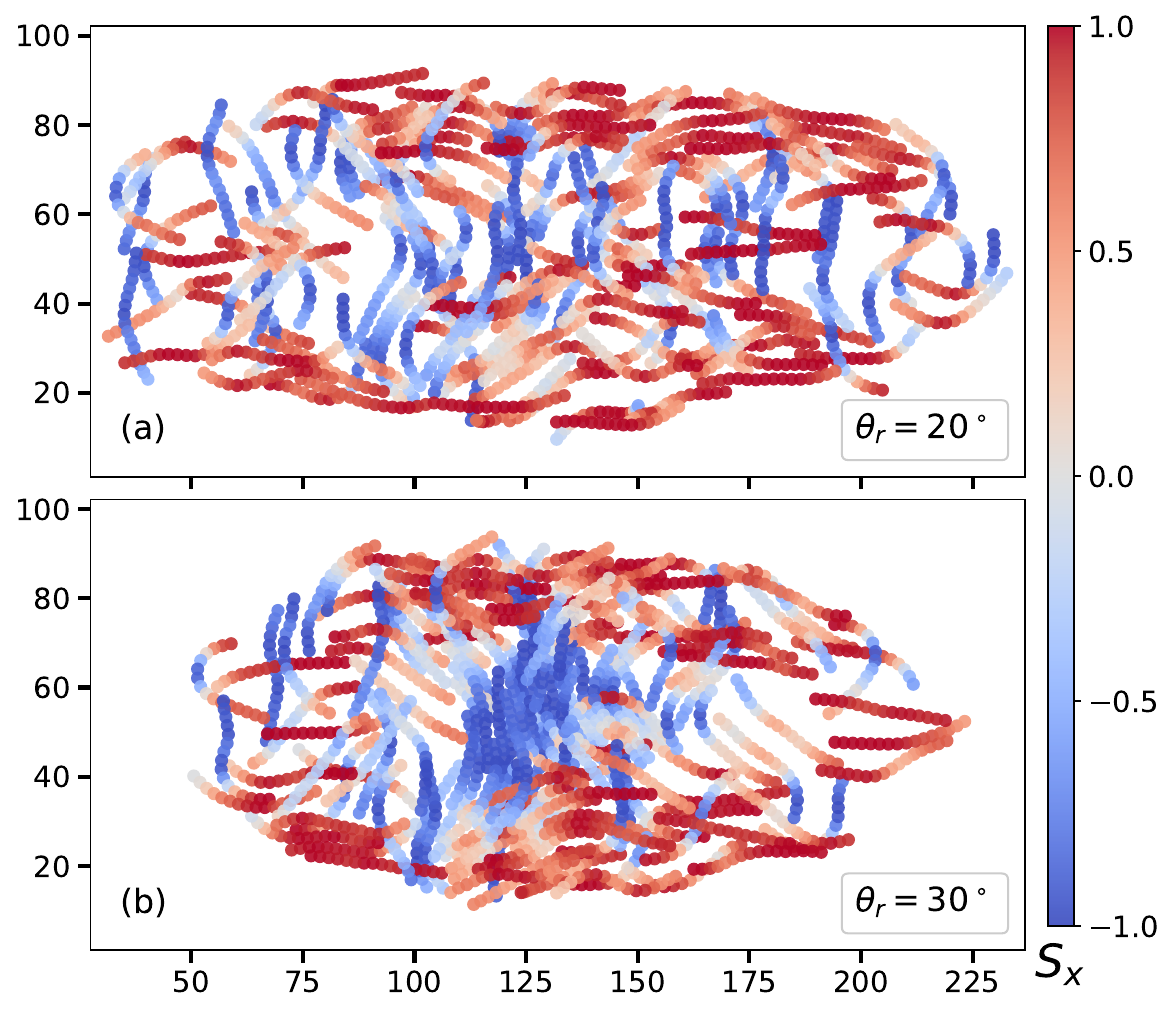}
    \caption{Segment-wise nematic order parameter of the final deposit for non-cohesive filaments ($L=32$) at receding contact angles a) $\theta_r = 20^\circ$ and b) $\theta_r = 30^\circ$. The order parameter is calculated between adjacent beads within each filament. For $\theta_r=20^\circ$, the deposit extends further axially because the contact line remains pinned longer; this allows stronger capillary flows to persist for a longer time, while the thinner film increases the probability that filaments come close to the substrate and thus experience substrate interaction and friction. For $\theta_r=30^\circ$, the contact line recedes earlier, pushing a larger population of filaments toward the central region, where they exhibit local ordering.}
    \label{fig:Nematic}
\end{figure}

\revisedtext{To isolate the effects of filament attraction, we first consider the filaments without cohesive interactions as a reference case.} We study the effect of the receding contact angle on the final deposition pattern. Fig.~\ref{fig:Nematic} shows top-view snapshots of the fully deposited non-cohesive filaments for receding contact angles of $\theta_r=20^\circ$ and $\theta_r=30^\circ$. The color scale represents the intramolecular nematic order $S_x$, where red ($S_x=1$) indicates axial alignment and blue ($S_x=-1$) transverse alignment. Increasing the receding contact angle to $\theta_r=30^\circ$ weakens capillary flows and promotes earlier depinning of the contact line, thereby reducing filament accumulation near the edge and allowing filaments to move more freely as the contact line recedes, since fewer filament segments anchor to the substrate and the effective friction is lower. In contrast, for $\theta_r=20^\circ$ the contact line remains pinned for a longer period, producing stronger capillary flows and a thinner liquid film. This leads to earlier adhesion of filaments to the substrate and results in an axial deposit that extends approximately $15\%$ further in the axial direction compared to the case with $\theta_r=30^\circ$.

\begin{figure}[h]
    \centering
    \includegraphics[width=0.99\linewidth]{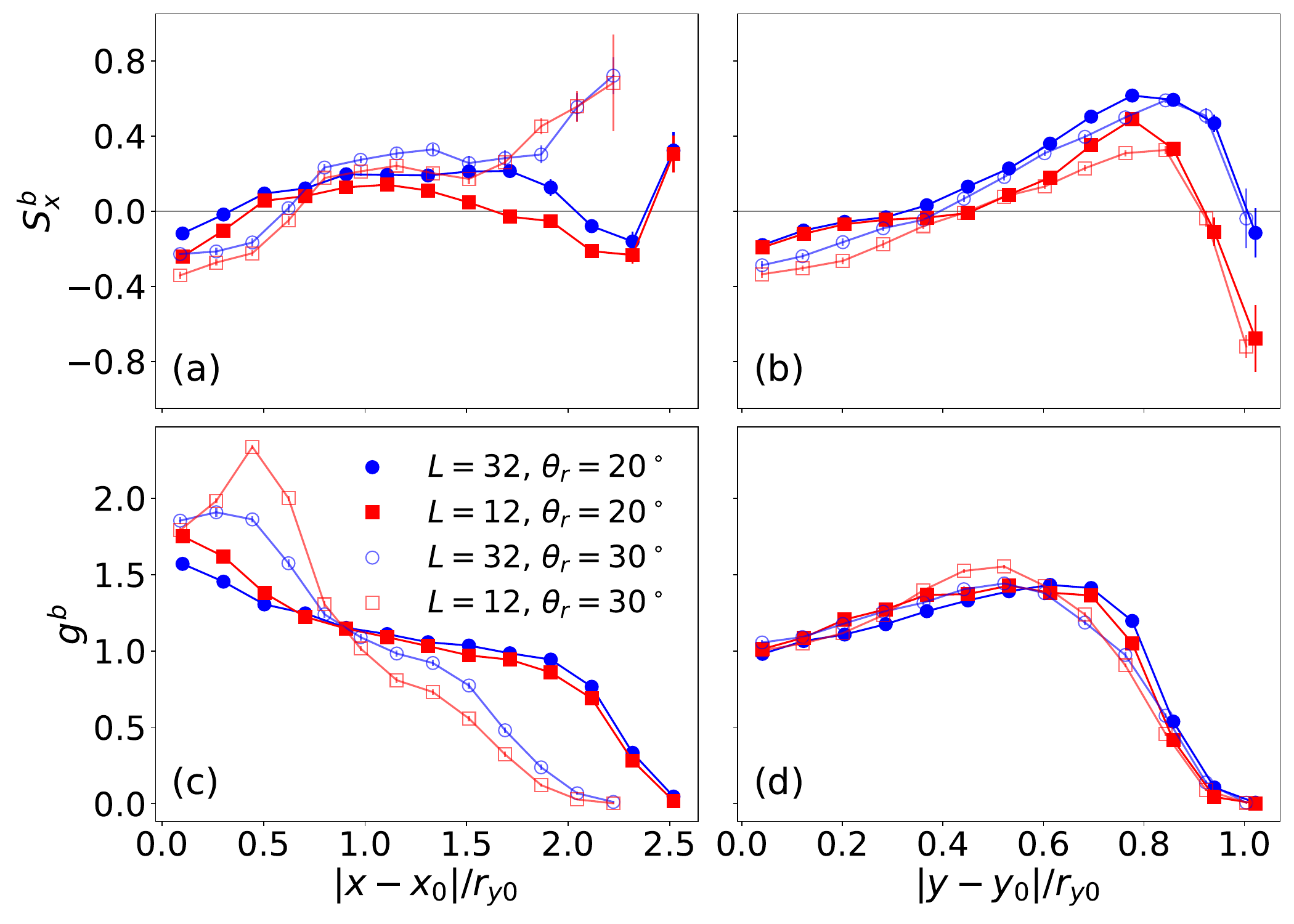}
    \caption{Top (a, b): Segment-wise nematic order $S_x^b$. Bottom (c, d): normalized density profile $g^b$. Profiles are shown in the axial ($x$, left) and transverse ($y$, right) directions for different filament lengths $L=12$ and $L=32$, and receding contact angles $\theta_r=20^\circ$ and $\theta_r=30^\circ$.
    \revisedtext{The quantities are evaluated in each spatial bin $b$ and averaged over independent simulation runs.} Larger receding contact angles lead to a shorter axial extent of the deposit and a more centrally concentrated density profile. In the transverse direction, the coffee-ring peak shifts slightly inward and axial alignment decreases. Longer filaments exhibit stronger axial alignment overall. Alignment is predominantly axial near the edges and more transverse in the center, with this contrast becoming more pronounced at higher receding contact angles.}
    \label{fig:Nematic2}
\end{figure}

In both cases, the transverse contact line remains pinned significantly longer than the axial one. Capillary flows therefore advect filaments toward this region, where they preferentially deposit and align along the edge, generating a high axial order. Afterward, the droplet reaches the shape of a spherical cap, then contracts fully in constant contact angle mode. During this final evaporation stage, the concentration of non-adhered filaments in the bulk is substantially higher for the larger receding contact angle. The increased bulk concentration enhances the likelihood of local ordering in the central region. Accordingly, for $\theta_r=30^\circ$ the central filament density is markedly higher and the orientation is predominantly transverse: many filaments initially align with the transverse capillary flows and are then stabilized in that direction during the final ordering process. This mechanism further amplifies the central filament density and reinforces alignment within the droplet interior.

Fig.~\ref{fig:Nematic2} quantifies the deposited morphologies by showing the nematic order $S_x^b$ of filament segments (top row) and the normalized density profile $g^b$ (bottom row) as functions of the distance from the droplet center.
\revisedtext{Profiles are plotted in the axial ($x$, left column) and transverse ($y$, right column) directions, with distances measured relative to the droplet center $(x_0,y_0)$ and normalized by the initial droplet radius in the transverse direction, $r_{y0}$.} Results are presented for long ($L=32$) and short ($L=12$) filaments at receding contact angles $\theta_r = 20^\circ$ and $30^\circ$. All profiles are averaged over 40 independent simulations and over their mirror images at the droplet center. \revisedtext{For each profile, the full deposit range, including the outer edges, is shown in the plotted direction, while averaging is restricted to the central half of the deposit in the perpendicular direction.}

In the axial direction, the smaller receding contact angle $\theta_r = 20^\circ$ results in an approximately 15\% larger deposit extent and a more pronounced edge accumulation, as shown in Fig.~\ref{fig:Nematic2}c. This reflects stronger contact-line pinning, which enhances capillary flows and promotes filament transport toward the contact line; the resulting accumulation increases the effective substrate friction, helping to retain filaments at the edge and hinder contact-line contraction. In contrast, the transverse deposit extent remains nearly unchanged with varying $\theta_r$, as the contact line in this direction remains pinned for a longer duration. Nevertheless, increasing $\theta_r$ leads to noticeable changes in the area-fraction profile (Fig.~\ref{fig:Nematic2}d): the coffee-ring peak shifts inward and the edge accumulation becomes less pronounced.
\revisedtext{Quantitatively, the peak shifts inward from $|y-y_0|/r_{y0} \approx 0.7$ for $\theta_r = 20^\circ$ to approximately $0.5$ for $\theta_r = 30^\circ$.}

\begin{figure*}[t]
    \centering
    \includegraphics[width=1.0\linewidth]{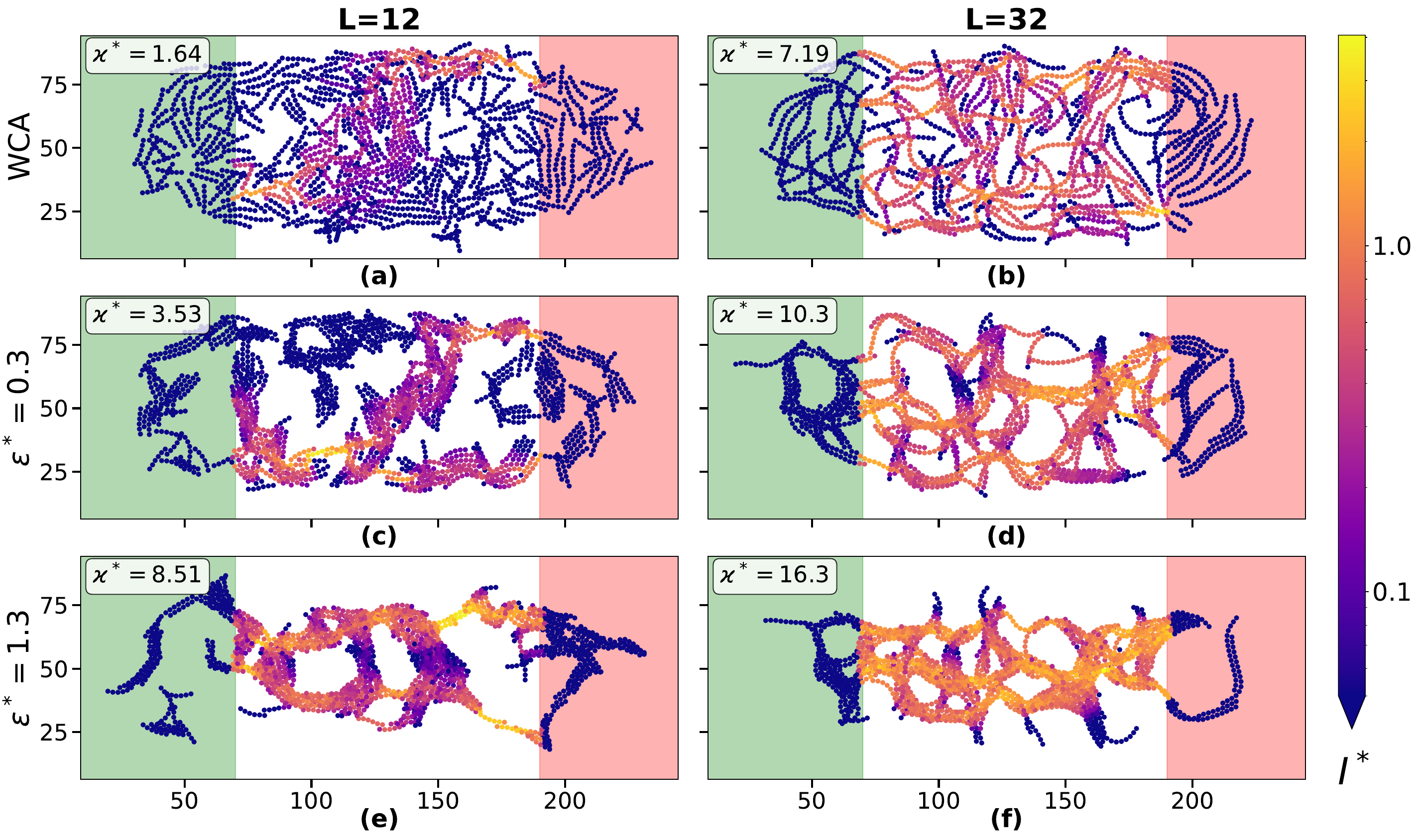}
    \caption{Final deposit morphologies at a receding contact angle of $\theta_r=20^\circ$. The first row (a, b) shows purely repulsive interactions (WCA potential), while the second (c, d) and third (e, f) rows correspond to attractive Lennard-Jones interactions with $\varepsilon^* = 0.3$ and $\varepsilon^* = 1.3$ respectively. Columns compare short ($L=12$, left (a, c, e)) and long ($L=32$, right (b, d, f)) filaments. The color scale represents the dimensionless current $I^*$, normalized by the current of a single straight filament directly connecting source to sink. Increasing cohesion promotes filament aggregation and enhances network connectivity. \revisedtext{Long filaments tend to form more uniform and interconnected pathways, whereas short filaments show more pronounced bottlenecks in the displayed configurations.} Source (green) and sink (red) regions are indicated.}
    \label{fig:Morphologies1}
\end{figure*}

Filament length strongly influences the orientational order. Longer filaments exhibit consistently enhanced axial alignment, as shown in Fig.~\ref{fig:Nematic2}b (blue symbols) compared to shorter filaments (red symbols). This can be attributed to their lower number density, which reduces filament-filament interactions, as well as to intra-filament alignment effects. These arise from shear-induced alignment or when a segment of a filament becomes anchored at the substrate or near the contact line; the remaining segments then tend to align with the local flow direction or the direction of the receding contact line. Under strong geometric confinement or higher local concentrations, this enhanced alignment can also promote more ordered packing of neighboring filaments.

Alignment is predominantly axial $S_x>0.5$ near the deposit edges, as shown in Fig.~\ref{fig:Nematic2} (top). This is most evident at the transverse coffee-ring positions (Fig.~\ref{fig:Nematic2}b) and along the axial boundaries (Fig.~\ref{fig:Nematic2}a). The axial ordering at the axial boundaries becomes more pronounced at larger receding contact angles $\theta_r=30^\circ$ (Fig.~\ref{fig:Nematic2}a, empty symbols), compared to $\theta_r=20^\circ$ (Fig.~\ref{fig:Nematic2}a, filled symbols), reflecting reduced contact-line pinning, enhanced contact-line motion, and consequently weaker coffee-ring accumulation. Along the axial boundaries (Fig.~\ref{fig:Nematic2}a), filament orientation is primarily guided by weak capillary flows and the direction of the receding contact line. In the transverse direction (Fig.~\ref{fig:Nematic2}b), filaments initially align with the contact line at the coffee-ring positions and subsequently with previously deposited filaments that have already adopted this axial orientation.

In contrast, the central region exhibits transverse alignment ($S_x<0.4$ in Fig.~\ref{fig:Nematic2}, top), which becomes more pronounced at larger receding contact angles ($\theta_r=30^\circ$, empty symbols). Here, weaker outward capillary flows, faster droplet contraction, and a less confined height profile near the contact line lead to higher filament concentrations during the late stages of evaporation. This is reflected by the higher normalized density $g^b$ in the central region (Fig.~\ref{fig:Nematic2}c). The resulting increase in local density promotes local ordering with predominantly transverse orientation. This orientation emerges because filaments are initially transported toward the transverse contact line, where they align with the local flow and contact line direction, and subsequently remain in this configuration as the droplet contracts inward.

\subsubsection{Conductivity}
Here, we relate the deposit morphology to the resulting relative electrical conductivity, focusing on the effects of filament cohesion and length. \revisedtext{
The results for filaments with non-cohesive interactions discussed above provide a reference case, while here we consider cohesive interactions that alter the deposit morphology and local ordering.} Fig.~\ref{fig:Morphologies1} shows the final deposition morphologies for different normalized filament cohesion strengths, $\varepsilon^* := \varepsilon / \varepsilon_s \in [0.3, 1.3]$, together with the purely repulsive reference case described by the WCA potential. Results are shown for short ($L=12$) and long ($L=32$) filaments at a receding contact angle of $\theta_r = 20^\circ$. The color scale represents the spatial distribution of the dimensionless current $I^*$ and thus highlights conductive pathways connecting source and sink. For each filament length, identical random seeds are used to generate comparable initial filament configurations. 
\revisedtext{Compared with the non-cohesive case, cohesion promotes aggregation and leads to distinct local filament arrangements. As the cohesive strength increases, the filaments progressively form larger aggregates and more pronounced structures, in qualitative agreement with previous work~\cite{Kröger2025}. 
The corresponding relative conductivity also increases over the explored parameter range, as shown in Fig.~\ref{fig:Morphologies1}c--f.} At the same time, the deposit becomes more spatially confined, predominantly in the transverse direction.
\revisedtext{Longer filaments tend to form more homogeneous and interconnected current-carrying pathways, consistent with our previous work~\cite{Schoettner2026}. This tendency can be attributed to their larger spatial extent, which allows them to bridge larger distances with fewer inter-filament contacts and therefore reduces the occurrence of local bottlenecks.}
Filament length strongly affects the resulting network topology. Longer filaments form more \revisedtext{extended} and better interconnected structures (Fig.~\ref{fig:Morphologies1}b,d,f), which increase the likelihood of uninterrupted source-sink pathways and lead to a more uniformly distributed current.
In contrast, short filaments (Fig.~\ref{fig:Morphologies1}a,c,e) produce more heterogeneous networks with larger depleted regions and pronounced transport bottlenecks. For weakly cohesive short filaments, this effect is further amplified, resulting in localized filament-rich clusters that remain disconnected or do not contribute to the main current-carrying backbone (Fig.~\ref{fig:Morphologies1}c). Aggregation initially leads to the formation of filament-rich regions, from which conductive pathways subsequently emerge. These pathways coincide with regions of high current and are further shaped by drying-induced density variations, such as the coffee-ring effect. Cohesion also promotes alignment within these regions, with filaments preferentially oriented along the emerging pathways (Fig.~\ref{fig:Morphologies1}c,d), an effect that is particularly pronounced for long filaments.
\revisedtext{The detailed deposition morphology can additionally depend on filament stiffness, which should be considered when comparing the present results with experimental nanowire or nanotube deposits.}

\begin{figure}[h]
    \centering
    \includegraphics[width=1.0\linewidth]{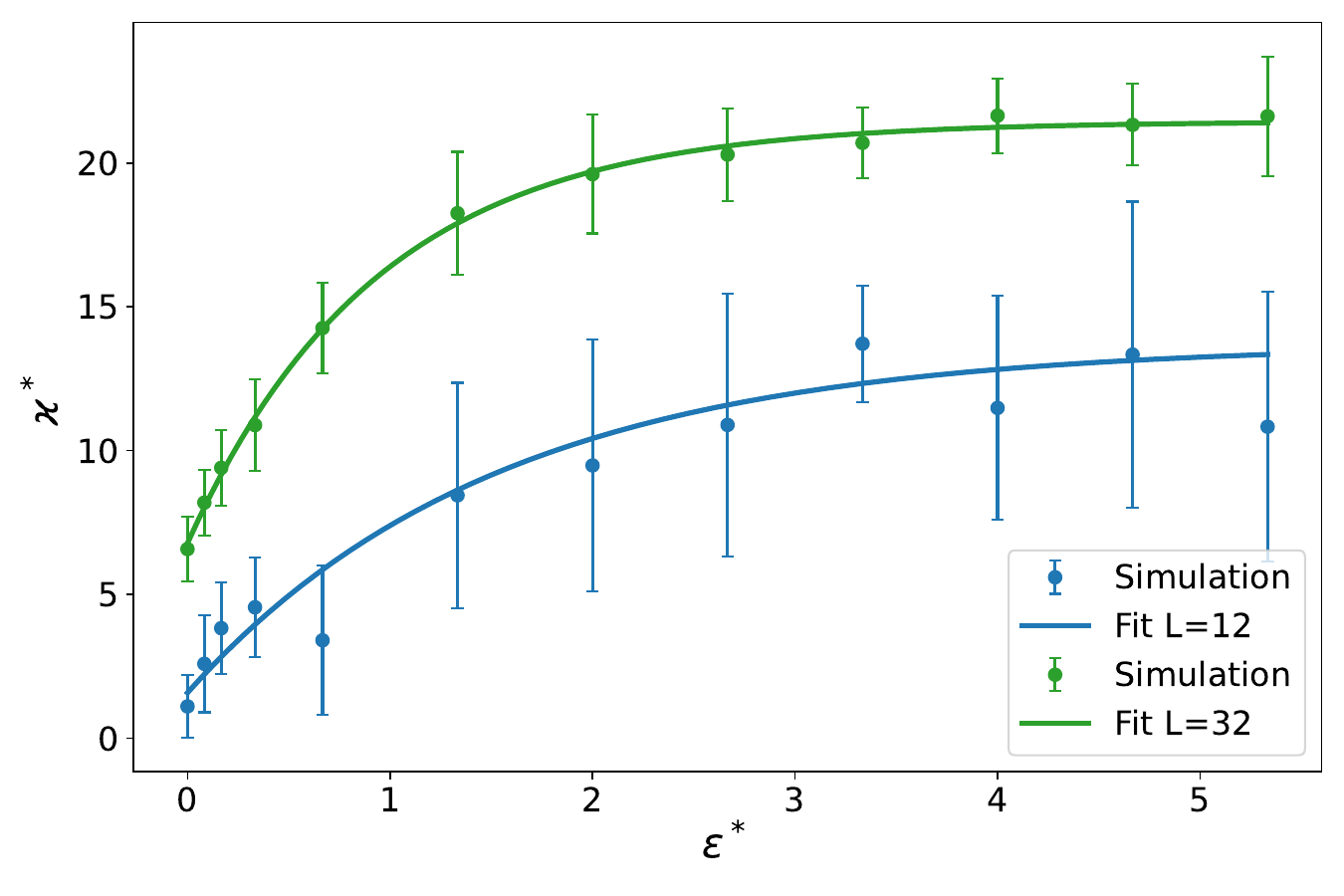}
    \caption{Relative electrical conductivity $\varkappa^*$ as a function of the normalized cohesion strength $\varepsilon^*$ for short ($L=12$) and long ($L=32$) filaments. The relative conductivity increases with both cohesion strength and filament length. Each data point represents an average over $20$ independent initial configurations.}
    \label{fig:CondVSAttract}
\end{figure}

Fig.~\ref{fig:CondVSAttract} shows the relative conductivity $\varkappa^*$ as a function of the normalized cohesive van der Waals interaction strength $\varepsilon^*$ for short ($L=12$, blue circles) and long ($L=32$, green circles) filaments. We note that the reported values of $\varkappa^*$ can depend on the chosen separation between source and sink electrodes, since the deposit is spatially non-uniform and strong cohesion can induce axial contraction. Here, the source-sink distance is fixed to 120 lattice units in all cases. Overall, increasing cohesion promotes filament agglomeration, which enhances network connectivity and thereby increases $\varkappa^*$. In addition, longer filaments consistently yield higher values of $\varkappa^*$.

The increase of $\varkappa^*$ with increasing cohesive interaction strength $\varepsilon^*$ shows a saturation-like behavior. This can be explained by the gradual formation of contacts between filaments. Each filament segment can only form a limited number of contacts with nearby filaments. When cohesion is increased, these contacts become more likely. At low cohesion, this can strongly increase the number of conductive pathways. At higher cohesion, many possible contacts have already formed, so that further increasing $\varepsilon^*$ has a weaker effect. This trend is described by a phenomenological exponential saturation model~\cite{Azizian2004}
\begin{equation}
\varkappa^*(\varepsilon^*) = \varkappa_0^* + (\varkappa_\infty^* - \varkappa_0^*) \left(1 - e^{-A \varepsilon^*} \right),
\end{equation}
which captures the rapid initial increase and the weaker growth at larger cohesion. We use this expression as a descriptive fit rather than as a microscopic transport theory. Here, $\varkappa_0^*$ represents the baseline conductivity of the non-cohesive network, $\varkappa_\infty^*$ the asymptotic fit value at large cohesion, and $A$ a parameter that depends on filament concentration, length, and interaction range. As shown in Fig.~\ref{fig:CondVSAttract}, this model (solid lines) qualitatively captures the dependence of $\varkappa^*$ on $\varepsilon^*$. Although no microscopic theory directly links cohesion to conductivity in filament networks, this minimal phenomenological model captures the essential statistical behavior: at low cohesion, many new contacts are formed for small increases in $\varepsilon^*$, whereas at high cohesion, most possible connections have already been realized, leading to saturation.

\section{Conclusion}

In this work, we investigated the evaporation-driven self-assembly of nanowire-laden elongated droplets on line-shaped hydrophilic patches using mesoscale simulations.
The nanowires are modeled as bead–spring filaments. By varying filament length, inter-filament cohesion, and substrate wettability, we linked drying dynamics to filament organization and the resulting electrical transport of the final deposit.

Our results identify geometry as the dominant organizing principle in elongated droplets. The intrinsic anisotropy of the droplet footprint imposes a two-stage drying pathway---initial axial contraction followed by radial recession---that produces pronounced axial and transverse inhomogeneities within the deposit. This geometry-imposed sequence strongly influences the flow fields, alignment patterns, and material redistribution, while microscopic filament parameters modulate these effects. Lower receding contact angles (e.g., $\theta_r = 20^\circ$) amplify the anisotropy by prolonging contact-line pinning, thereby enhancing edge-directed flow and local coffee-ring accumulation. The resulting deposits are more elongated and exhibit increased nematic alignment arising from shear flows and geometric alignment along the contact line. In contrast, earlier depinning ($\theta_r = 30^\circ$) promotes central material accumulation and local ordering, leading to reduced axial extension and weaker global alignment.

Beyond geometric effects, we disentangle the distinct roles of inter-filament cohesion and filament length. Increasing inter-filament cohesion enhances the relative conductivity by promoting aggregation and inter-particle contact formation. However, this improvement comes at the cost of structural homogeneity, as strong cohesion promotes clustering and the formation of dense, spatially heterogeneous domains. In contrast, increasing filament length provides a dual benefit: longer filaments strengthen long-range connectivity through more stable percolating backbones while simultaneously suppressing excessive local densification, resulting in more spatially uniform conductive networks. 
\revisedtext{This interpretation is supported by the density profiles, which indicate a slightly more uniform spatial distribution for the longer filaments, and by the consistently higher average conductivity observed for longer filaments over the investigated range of cohesion strengths.}

Taken together, our results highlight three main control parameters in evaporative line formation: geometric confinement, cohesive interactions, and filament aspect ratio. Within the studied parameter window, geometric confinement sets the global deposition pathway and spatial anisotropy. Cohesive interactions govern local packing by increasing inter-filament contacts, thereby improving relative conductivity but also promoting clustering and structural heterogeneity. In contrast, filament aspect ratio determines network-scale connectivity: longer filaments stabilize percolating backbones, improving relative conductivity while maintaining structural uniformity. These findings provide a useful design perspective in which substrate wettability controls macroscopic deposition pathways, inter-filament cohesion tunes local contact formation, and filament length affects global percolation efficiency. By adjusting these parameters, filament alignment and the trade-off between relative conductivity and uniformity can be systematically tuned, which is relevant for the design of filament-based deposits in printed electronics~\cite{Goh2019}.

\section*{Author contributions}
J. Sch\"ottner: conceptualization, methodology, software, formal analysis, investigation, visualization, and writing--original draft. Q. Xie: methodology, software, supervision, and writing--review and editing. J. Harting: conceptualization, supervision, funding acquisition, and writing--review and editing.

\section*{Conflicts of interest}
There are no conflicts to declare.

\section*{Data availability}
The data that support the findings of this study are openly available at \url{https://doi.org/10.5281/zenodo.21242492}.

\section*{Acknowledgements}
We thank Johannes Hielscher and Johannes Beunen for supporting the development of the simulation code. We acknowledge financial support from the Deutsche Forschungsgemeinschaft (DFG, German Research Foundation) -- Project-ID 528402728 (research group ``3D-HF-MID'') and Project-ID 416229255 (CRC 1411 ``Design of Particulate Products''). Further financial support is acknowledged from the German Federal Ministry of Education and Research (BMBF) -- Project H2Giga/AEM-Direkt (grant number 03HY103HF). We thank the Gauss Centre for Supercomputing e.V. (\url{www.gauss-centre.eu}) for funding this project by providing computing time through the John von Neumann Institute for Computing (NIC) on the GCS Supercomputer JUWELS at J\"ulich Supercomputing Centre (JSC).

\begin{comment}
\section*{Acknowledgments}
We thank Johannes Hielscher and Johannes Beunen for supporting the development of the simulation code. 
We acknowledge financial support from the Deutsche
Forschungsgemeinschaft (DFG, German Research Foundation) -- Project-ID 528402728 (research group ``3D-HF-MID") 
and the German Federal Ministry of Education and Research (BMBF) -- Project H2Giga/AEM-Direkt (Grant number 03HY103HF).
We thank the Gauss Centre for Supercomputing e.V.(\url{www.gauss-centre.eu}) for funding this project by providing computing time
through the John von Neumann Institute for Computing (NIC) on the GCS Supercomputer JUWELS at Jülich Supercomputing Centre (JSC).
\section*{Declaration of competing interest}
The authors declare that they have no known competing financial interests or personal relationships that could have appeared to influence the work reported in this paper.
\section*{Data availability}
The data that support the findings of this study are openly
available at \url{https://doi.org/10.5281/zenodo.21242492}.
\end{comment}

\bibliography{Bibfile}

\end{document}